# DesignCon 2021

# Capacitor Optimization in Power Distribution Networks Using Numerical Computation Techniques

Jordan R. Keuseman, Mayo Clinic
keuseman.jordan@mayo.edu

Chad M. Smutzer, Mayo Clinic
smutzer.chad@mayo.edu

Clifton R. Haider, Mayo Clinic
haider.clifton@mayo.edu

Barry K. Gilbert, Mayo Clinic
gilbert.barry@mayo.edu, 507-284-4056

## Abstract

This paper presents a power distribution network (PDN) decoupling capacitor optimization application with three primary goals: reduction of solution times for large networks, development of flexible network scoring routines, and a concentration strictly on achieving the best network performance. Example optimizations are performed using broadband models of a printed circuit board (PCB), a chip-package, on-die networks, and candidate capacitors. A novel worst-case time-domain optimization technique is presented as an alternative to the traditional frequency-domain approach. The trade-offs and criteria for scoring the computed network are presented. The output is a recommended set of capacitors which can then be applied to the product design.

## Author Biographies

**Jordan R. Keuseman** received a BS in Computer Engineering from Minnesota State University (Mankato, MN). He is currently a Senior Engineer at the Mayo Clinic Special Purpose Processor Development Group where his focus has been power integrity and power electronics.

**Chad M. Smutzer** received a BSEE from the University of Iowa in Iowa City. He is currently a Senior Engineer at the Mayo Clinic Special Purpose Processor Development Group, where he performs signal and power integrity analysis.

**Clifton R. Haider** received a BS in Biomedical Engineering from the University of Iowa (Iowa City, IA) and a Ph.D. in Biomedical Engineering from Mayo Clinic (Rochester, MN). He is currently Deputy Director of the Mayo Clinic Special Purpose Processor Development Group, directing research efforts in high-performance electronics and related areas.

**Barry K. Gilbert** received a BSEE from Purdue University (West Lafayette, IN) and a Ph.D. in physiology and biophysics from the University of Minnesota (Minneapolis, MN). He is currently Director of the Mayo Clinic Special Purpose Processor Development Group, directing research efforts in high-performance electronics and related areas.

# I. Introduction

Power distribution networks (PDNs) for high-performance digital systems involve careful design considerations from the voltage regulator (VR) to the load device (ASIC, FPGA, CPU, etc.). Without a properly engineered power delivery system a load device can fail to operate reliably. When the load malfunctions the power supply is often investigated first as a potential explanation. Not only is the static voltage important but PDN immunity to fluctuations from sudden load demands is also very relevant. An essential facet to high-quality power integrity (PI) design is selecting the best network of capacitors to create the desired impedance profile from the VR to the load device. A PDN with poorly chosen capacitors may experience transient droops or overshoots that can cause errors or even damage the load devices. The PDN response can vary based on the workloads and resulting load patterns. With shrinking transistor feature sizes in modern load devices, the power requirements trend toward lower voltages and higher currents. In addition, the load voltage tolerance specifications rarely are given any relief to combat these scaling trends, which pushes the PI design toward more elaborate designs and technologies.

For many low-cost designs, the challenge is to find the minimum number and type of capacitors to provide adequate margin for the load to operate. Frequently the design has marginal corner conditions which are ignored because low-probability failures are considered acceptable. A high-reliability product may consider any failure to be unacceptable, and therefore margin must be maintained. For high-performance designs, the key metric is often the overall system performance requirements; however, power delivery is frequently the limiting factor, sometimes caused by physical-space limitations. The key focal-point for this paper is the optimization of PDN decoupling capacitor networks for high-performance and high-reliability systems where performance and robustness are the primary considerations.

When working with complex PDNs most engineers rely on simulation software to assist in the design. Many commercial electromagnetic (EM) software packages exist and the PI analysis capabilities continue to evolve and improve. Available software now offers PDN capacitor network optimization for target impedance, and some even allow control over the optimization goal (cost, quantity, etc.). Presently, the market demands tend to focus on achieving adequate performance at a minimal cost, and industry tools accommodate those scenarios. For high-performance and/or high-reliability products the features that these tools provide are not always adequate or able to be extended in new ways. In particular, we were analyzing complex designs with over 500 capacitors and the goal was to achieve performance levels that were at the limit of the PDN technology. The physical space limitations were such that every component had to be chosen to optimize performance. In these design classes the goal was not to minimize the number of capacitors, but to maximize their effectiveness.

The performance of a PDN can be viewed as impedance versus frequency, or Z(f). PI engineers often use this visualization technique to compare to a required target impedance curve as validation of the design performance [1]. Using the target impedance methodology, an engineer can very quickly assess the PDN's performance across a selected frequency band. The methodology to calculate the target impedance curve will not be discussed in detail here. However, in recent years several papers have been published on the perils of only considering target impedance in a PDN design [2]. The "rogue waves" concept comes from oceanographers studying the effects of constructive interference of ocean waves, causing extremely large or dangerous waves to form. This rogue wave phenomenon is analogous to the resonances which can form in a PDN, and can cause worst-case voltage waveforms with load patterns that can stimulate these anti-resonance peaks [3]. Achieving the target impedance is only part of the equation, while the other part is ensuring that resonances do not cause transient voltage violations. The most robust capacitor network should be designed to maintain relatively flat impedance profiles to avoid significantly large or high-Q anti-resonant peaks [4].

Time-domain analysis is often performed using EM-aware simulations, or with SPICE modeling to characterize step-load behavior, for example. This analysis is typically conducted after some initial engineering is performed to design the PDN; included in that simulation are the decoupling capacitors and may include the VR and load stimulus models. The output of this analysis is useful, but difficult to evaluate across widely varying workload patterns such as those that might be required to assess rogue waves. The impedance profile of a PDN can significantly alter its time-domain behavior, and change the necessary patterns required to obtain worst-case transient noise. It is usually beneficial to employ frequency-domain *and* time-domain simulations to gain a broader picture of the performance characteristics.

To further complicate the PDN engineering efforts, there are different goals and philosophies, often dictated by the specific load device being powered. Power management logic integrated into the load is finding increasing use, in order to maintain stable operation amidst transient noise that is difficult to fully characterize during the engineering and test phases of product development [5] [6]. Although these technologies show promise, they are no panacea. However, these technologies can define a unique prioritization between the goals of having ultra-low impedance, coupled with flat and stable operation. For example, it may be advantageous in some designs to minimize a region of narrow-band impedance at the expense of anti-resonant behavior in some other band. There is not always a universally optimal design.

The standard PDN optimization methodology utilizing target impedance is useful, but is not always an ideal scoring metric when comparing alternative capacitor schemes. Even if a design meets the target impedance goal, it is not easy to decipher from the Z(f) results which PDN would give the lowest power rail noise when confronted with a step-load [7]. For this reason, we developed a time-domain worst-case transient response optimization routine as a new way of obtaining the best performance with a clear, quantifiable score. To support the wide range in optimization goals the application that we developed for this paper offers several ways of scoring a solution, in order to help the engineer assess the most applicable performance metric. Our application allows for an exploration environment, where new optimization ideas can be experimented with by adding a new routine.

# II. Application Structure

The capacitor optimizer application that we developed had the following goals to enhance existing solutions:

- Scalable to large PDN structures
- Able to analyze a complete PDN by combining models for the board, package, chip, VR, etc.
- Flexibility to create new PDN scoring routines for specific optimization goals
- Incorporating time-domain analysis into the optimization

The application that we constructed to fulfill these requirements used the MATLAB programming and computing environment. Figure 1 is intended to clarify the key user inputs, generated outputs and libraries used for the application, each of which will be described in the sections to follow.

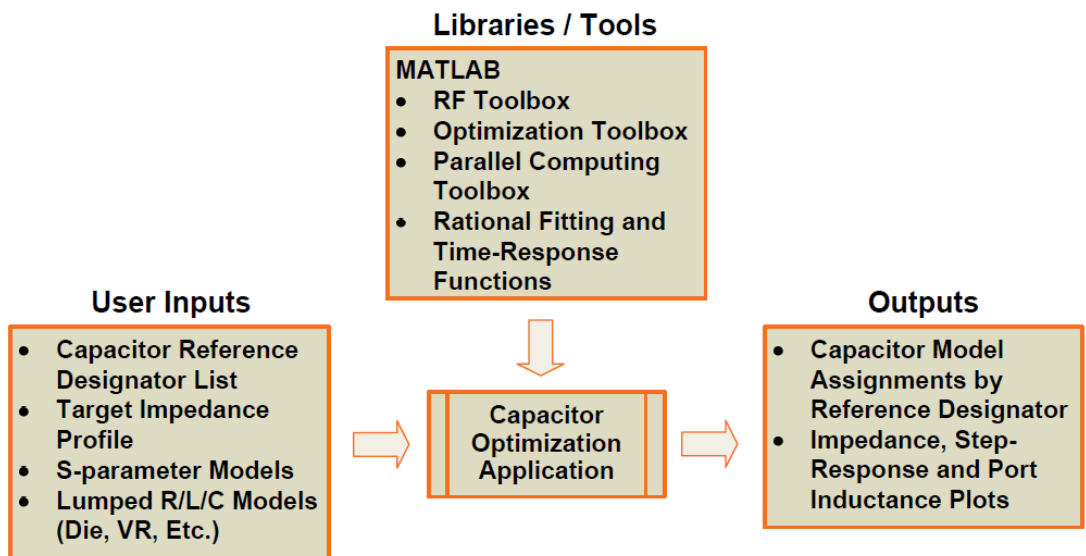

Figure 1 – Summary of application inputs, outputs and libraries used (46628)

## User Inputs

We took advantage of commercially available EM tools to extract models of complex PCBs and package substrates for analysis. These EM modeling tools produced frequency-domain S-parameter models of the layout artwork, which we exported in the industry-standard Touchstone format for external use. Our optimizer application makes no effort to control the physical placement of the chosen capacitor sites; industry tools already do an adequate job of helping to identify poorly performing capacitor sites, as well as driving placement of capacitors to more favorable locations. Our recent experiences suggest that many high-performance designs have limited and very specific areas defined for decoupling capacitors, and existing tools are adequate to control the physical placement of those parts. A requirement for our capacitor optimizer application is to assign ports within the model at each candidate capacitor site, which will be used later in the optimization process to identify the best capacitors to be attached at each location.

Many high-performing PDNs involve a PCB attached to an IC substrate package, through some passive interconnect. Some EM tools possess the ability to physically merge the PCB and IC substrate designs together with a model of the interface to obtain a singular electrical model. This merging operation allows more precise and distributed modeling of the intricate power delivery from the board through the package and to a die. If IC models are available, these too can be incorporated into the stack of models. If detailed IC models are not available, a simple lumped model of on-die capacitance is usually sufficient to represent the critical anti-resonant peaking due to the interaction of the PDN inductance with the on-die capacitance.

In order to achieve the best PDN modeling accuracy, candidate capacitor models must also be obtained in Touchstone format, either through the capacitor vendor's website, or through measurements performed with a vector network analyzer. It is critical to obtain capacitor models for the specific package sizes allowable within the design constraints. Multi-layer ceramic capacitors have strong correlations between the equivalent series resistance (ESR) and equivalent series inductance (ESL) depending on the package size. In general, a larger package for a given capacitance will have lower ESR, but higher ESL [8]. Typically, designs greatly benefit from reductions in inductance from smaller packages, but controlling ESR can be a critical factor as well.

A VR model can often be helpful if low-frequency behavior is required; such a model can take many forms with the most simplistic model consisting of a series resistor and inductor [9]. Usually, a VR model is only required if time-domain analysis is performed or if the target impedance extends down to frequencies where the VR can influence the impedance, i.e. if the VR control-loop bandwidth intersects or approaches the frequency band of the target impedance.

The target impedance information used in the frequency-domain scoring is defined in a simple text file with frequency/value pairs. Our application generates the target impedance curve by connecting these frequency/value pairs together, using a log-frequency scale to ensure that straight lines are generated with the traditional log-log impedance plots. There is no limit to the target impedance curve complexity, but at least two pairs of data points must be specified.

The list of candidate capacitor models are simply supplied by a text file, which can contain any number of rows of capacitor models. Our application uses the filename in the text file to process the capacitor models into MATLAB S-parameter models.

## Libraries and Tools

The RF Toolbox incorporated into MATLAB provides a number of useful features that we took advantage of for working with network parameters. We were able to use this toolbox to read and write S-parameter files and perform necessary conversions between S, Y, Z or ABCD parameters. This toolbox also works well with the rational function fitting procedures and time-step functions that we will discuss later in the paper. We made some extensions to the default toolbox feature-set in order to scale to large PDN structures and perform efficient N-port model cascading and port reductions. The details of those extensions will not be discussed.

MATLAB's Optimization Toolbox contains a plethora of functions to minimize or maximize objectives. We investigated each of the available optimizers during the development of the application in order to find the most suitable one. The Parallel Computing Toolbox was well-integrated into the optimization routines and allowed us to take advantage of modern multi-core processors to improve the solution computation time.

## Outputs

The capacitor optimizer will provide plots of the optimized network response, but the critical output is a list of capacitor model assignments. The assignments can be used to adjust the PDN capacitors to the computed optimal set.

## Capacitor Optimizer Application Components

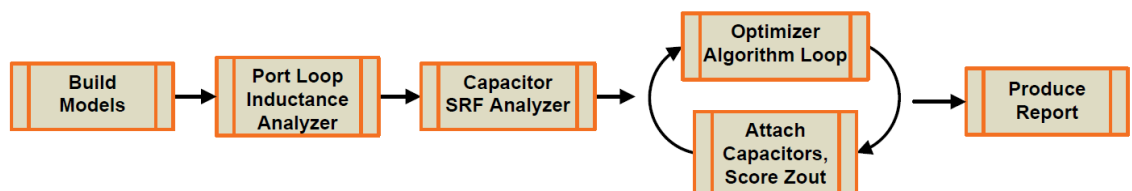

Figure 2 – Application structure flowchart (46589)

As diagrammed in Figure 2, the capacitor optimizer application begins by importing Touchstone S-parameter models for all the components and constructing MATLAB RF models. Then the application computes the loop inductances for each candidate capacitor site using the S-parameter model of the circuit board. The series resonance frequency (SRF) analyzer determines the resonance frequency for each of the candidate capacitor models. Our application then loads the models into the optimization routines, which attaches possible capacitors to each capacitor port, scores the network, and converges on an optimal solution to be presented to the user. Each of these components will now be discussed in more detail.

## Build Models

The PDN S-parameter model must contain an optimization observation port, typically the load device, and ports for each capacitor site to be included in the optimization solution. For our experiments we configured the load device pins in parallel with each other to simulate the effective impedance of the parallel current path through the load interconnect. Other ports such as for a VR or die model can be included as well.

Optionally, circuits can be built using discrete resistors, capacitors and inductors and connected to the model as applicable. The discrete components are converted to S-parameter equivalent models, which can be a useful way to represent lumped on-die models when broadband IC models are unavailable.

## Port Loop Inductance Analyzer

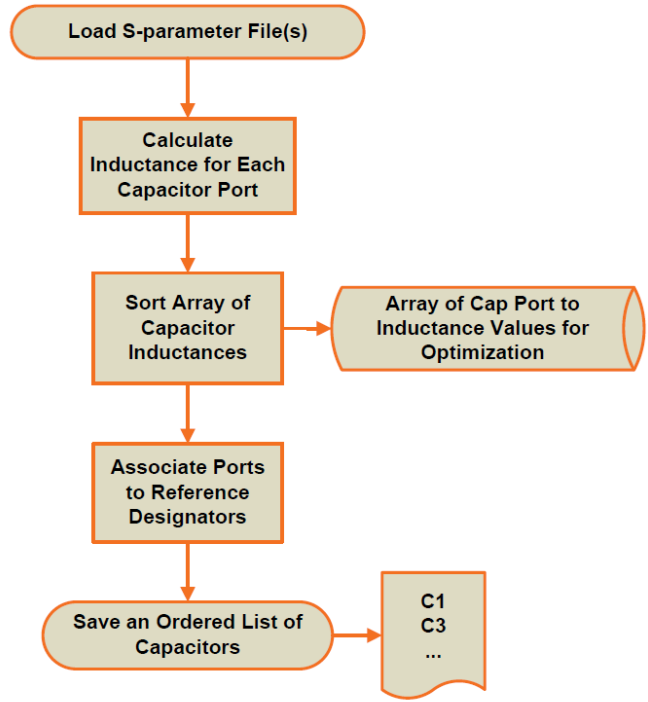

**Figure 3 - Loop inductance analyzer flowchart (46590)**

Capacitors that have low inductances between their mounting location and the load will perform the best. It can be difficult to determine visually which capacitor sites have the lowest inductance without analysis. The inductance analyzer routine iterates through each candidate capacitor port and determines the loop inductance, as illustrated in Figure 3. First we attach a short at the observation port and generate Z-parameters for each port location. The complex impedance is used to determine the effective inductance using the well-known relationship:

$$L = \frac{Z_{imaginary}(f)}{2 \times \pi \times f} \qquad (1)$$

The frequency (f) is chosen to be sufficiently high for the inductance to contribute quantifiable impedance (or reactance). Additionally, our application outputs the sorted list of capacitor reference designators into a text file useful for identifying the best and worst performing capacitor sites.

## Capacitor SRF Analyzer

The SRF for a capacitor is the frequency at which a resonance is formed by the intrinsic capacitance and packaging inductance. Our application has no way of knowing this frequency from the S-parameter information alone and must compute it. We perform the calculation without regard to the mounting inductance simply by converting the S-parameter data to Z-parameters and then locating the minimum impedance frequency. The SRF is stored for each candidate capacitor.

## Optimizer

A critical component of our application is the optimization method. PDN optimization is treated as a minimization problem, where the parameter to be minimized could be the Z(f) of the network or the transient response to a step-load. We wrote a function which abstracts the details of what constitutes a good PDN and simply outputs a numerical score based on a set of candidate capacitors. The independent variables define the specific quantities of the capacitor models ($Q_N$) to be attached to the N-port S-parameter model:

$$score = PDN.attach\_and\_score(Q_1, Q_2, \dots Q_N) \tag{2}$$

The optimizer method will iteratively adjust the quantities of each capacitor type and use our custom scoring routines to determine the effectiveness of that set. A number of optimization methods are available in the Optimization Toolbox in MATLAB, and the selection is not straightforward. Most classical optimization functions will produce a numerical output given a set of independent variables, but the variables are not limited to integer quantities. Of course a capacitor cannot be fractional. This integer optimization constraint eliminates a large number of the available optimizers. We caution the user who attempts a naïve fractional truncation that can cause improper solution convergence. Capacitor optimization contains a large number of variables to solve and the variables must have the linear algebraic constraint:

$$\boldsymbol{Capacitor\ Ports} \geq \sum_{i=1}^{N} \boldsymbol{Q_N} \tag{3}$$

Optimization problems may produce a single global minimum or potentially several local minima and a global minimum. For PDN networks, it is possible to begin converging to a local minimum without a broad search routine to decipher if this minimum is the global one. Genetic Algorithms (GA) are well-suited for PDN optimization problems because they can be adjusted to spread widely across the solution space, helping uncover the global minimum, and solving integer programming problems with linear algebraic constraints. Our application uses the GA to optimize the PDN with the ability to change to another algorithm if desired.

## Scoring

One of the most challenging aspects of the optimization problem is determining what constitutes ideal behavior. We would argue that there is no universal answer to an optimal PDN, but having low impedance, coupled with a flat response over frequency, is a general goal. We created a number of scoring concepts such as:

- Maximum Z(f) peak impedance above the target impedance curve
- Integration of the area between the Z(f) curve and the target impedance curve
- Flatness of Z(f)
- Transient step-response

All of the scoring concepts were implemented with modifiable weighting options to compute the final score for the optimizer. Additional schemes could easily be added for future enhancement. Although a perfectly flat impedance response is highly desirable, in practice it is very hard to achieve for ultra-low impedance PDNs. The step-response or time-domain approach we found highly desirable because it computes a single, quantifiable result (mV per ampere of current) that is easily compared across candidate solutions. The transient results that our application can generate should help address the desires for a flat response, and achieving target broadband impedance.

# III. Implementation Details

## Optimizer - Genetic Algorithm Tuning

For the reader who is unfamiliar with genetic algorithms, the basic premise is to emulate the natural selection process found in nature. A genetic algorithm (GA) will start by creating a randomized population of candidate solutions. The population is then scored and evolves into a new child population. Selection, crossover and mutations rules are used to generate child populations that try to maintain the best properties of parents while ensuring enough variation to formulate a wide search. The specific GA in the Optimization Toolkit is implemented with the Augmented Lagrangian Genetic Algorithm (ALGA) but when integer constraints are involved the penalty, crossover, mutation and fitness functions are modified.

Many adjustments can be made to the GA to speed convergence, at the expense of breadth of search. We found that the following settings were reasonable starting points as a compromise between speed and thoroughness:

- Maximum number of population generations to min ( $10 \cdot Cap_{types}$, 100 ) where $Cap_{types}$ is the quantity of capacitor types being considered
- Population size to min ( $\max(10 \cdot Cap_{types}, 40)$, 100 )
- Crossover fraction to 0.75
- Elite count to max( floor( $Cap_{types} \cdot 0.2$ ), 2 )
- Function tolerance to 0.001

## Attaching and Scoring

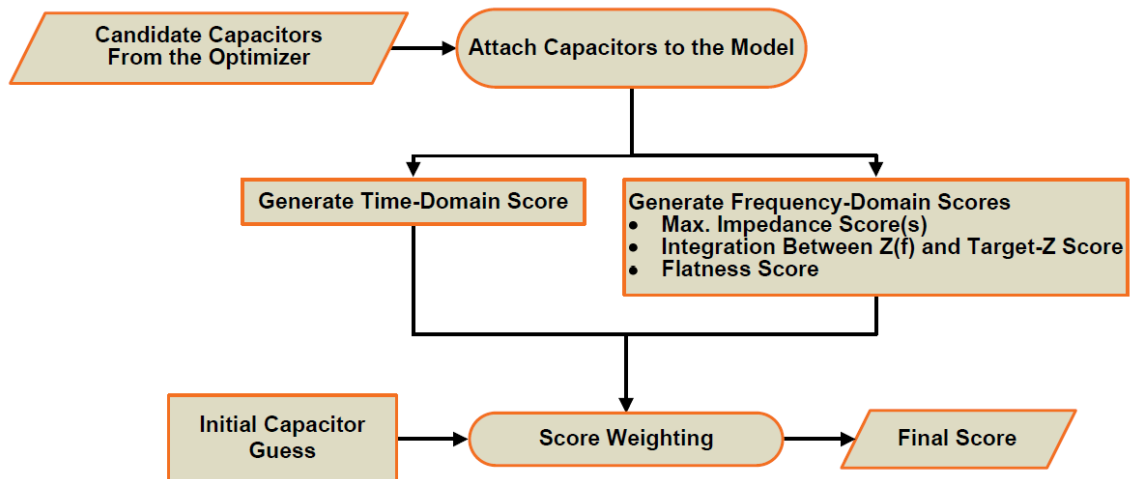

**Figure 4 – Capacitor attachment and scoring flowchart (46591)**

The GA optimization routine passes an array of capacitor model quantities to the attach routine. The attach routine iterates through this array and determines which specific capacitor port location should receive a specific capacitor model. The capacitor port inductance and SRF arrays that were previously calculated come into use at this point, as our routine determines the most effective location for a given capacitor type. Our application can be configured to attach capacitor models preferentially, based on SRF, by descending or ascending port inductances. The default behavior is to attach high-SRF capacitors to low-inductance ports.

The specific methodology that we used to attach capacitor S-parameter models to the PDN model is considered a port reduction operation, whereby the resulting model will combine the capacitor model and then remove that port from the model. The resulting model is then N-1 ports for each subsequent attached capacitor. The routine that we used leverages well-documented multi-port network connection routines [10].

After the candidate capacitors are attached, the Z-parameters for the observation port are computed. The Z(f) is used for scoring the individual solutions a number of ways as indicated in Figure 4**Error! Reference source not found.**. The scores from each of our methods are weighted and summed according to the user's preferences, and the final score is returned to the optimization routine.

Details on generating the initial capacitor population guess and the frequency-domain and time-domain scoring methods that we implemented will now be discussed.

## Initial Capacitor Population Guess

The total output score can be a combination from several scoring methods, which leads to a scaling complication in summing to the final score. A capacitor population may result in a widely varying score from one particular method that causes other scoring criteria to be neglected when summing to the total score. A normalization procedure must be applied to each individual scoring result. In order to determine the basis for the normalization, we used a simple heuristic to arrive at an initial capacitor population guess. The guess is then scored for each individual scoring metric and the value is stored for later comparison. Subsequently, when a new network is scored, our application recalls the initial guess and computes the ratio of the new solution to the original solution. A value of one would indicate that the current network is identical in performance to the initial guess. This procedure results in less variation between scoring criteria and each result is then scaled more simplistically for the final combined score.

We computed the initial capacitor guess by moving along the target impedance curve from minimum to maximum, and making a capacitor voting decision at each frequency step. At each frequency interval, the capacitor that has the closest SRF for the current frequency is given a scoring point to be used later in the population guess. After all the frequency intervals have been analyzed, the available capacitor sites are distributed based on the accumulated scoring points given to each capacitor type. This method gives no consideration to the mounting inductance of the capacitor ports or interaction between capacitors. Although simplistic, it is usually sufficient in generating a reasonable guess to establish a baseline score.

## Frequency Domain Scoring

Generally, keeping PDN impedance profile peaks below the target impedance is recommended, but not always practical. The next logical approach is to minimize the excursions above the target impedance. It is not reasonable simply to integrate the area under $Z(f)$, as that approach would not take into account the magnitude above or below the target impedance. We separately integrated the areas of impedance above and below the target impedance curve to serve as a scoring metric using equal frequency weighting. The area below the target impedance is only used if credit should be given to solutions that over-achieve in various regions. Many designers may not want to give credit for over-achieving in certain regions while violating the target in others, and want to concentrate on ensuring that the target is maintained across a wide bandwidth. Other designers might be less concerned about mild violations of the target impedance but very concerned with gross violations, so our application also supports scoring on maximum violations, as a function of distance from the target impedance.

There is a body of literature [4] [7] [11] that advocates for a PDN with a flat $Z(f)$ response. Flatness is easy to visualize but not as straightforward to compute. Flatness can be considered as limiting the deviation from the average impedance of the network. Our application can compute the average PDN impedance (across the target impedance bandwidth) followed by a summation of the difference between the broadband $Z(f)$ and the computed average. Alternatively, flatness can also be considered as limiting the anti-resonance (or resonance) Q-factor of each peak in $Z(f)$. For this reason, another scoring method that we created is to search $Z(f)$ and find local minima and maxima, which define these resonance/anti-resonance peaks. We then compute the Q-factor of the minima/maxima and sum to the total. This computation results in solutions that favor less sharp peaks (lower Q-factor) and reduced quantity of peaks.

## Transient Domain Scoring

To generate an estimate of the step-load transient voltage response, our application fits a rational function to the frequency response profile of each candidate solution. Since PDN analysis is typically based on Z-parameters, it made sense for us to use the Z-parameter data to perform the fit so that the results obtain the best accuracy. The rational fitting procedure [12] results in a Laplace transfer function equivalent model for the given Z-parameter data at the discrete frequency points. The fitting function constrains the minimum to maximum number of poles, which helps ensure that the model achieves a sufficient amount of complexity to model the full PDN behavior based on the number of resonance/anti-resonance characteristics. A range of 10 to 50 poles seemed sufficient to model most PDN Z(f) curves that we attempted. To ensure that the critical peaks were captured, we applied a per-frequency rational fit weighting. The weighting was set to the absolute value of the impedance. More frequency points ensure a better fit and more accurate results, at the expense of processing time. Rational functions are inherently causal and passivity can be inspected and enforced if necessary.

The rational function is then used by our application to compute the time response, given either as an impulse or step function. For PDN analysis, the load typically is defined by a maximum slew-rate ($\Delta i/\Delta t$), or a minimum rise-time. The minimum load rise-time is bounded by the ramp rate of current in the load circuits, and/or the series inductance between the PDN optimization port and the load. Once a load rise-time is determined, then it is possible to determine a more realistic PDN step-response than an impulse-response would compute. From Fourier analysis, a pulse with a finite rise-time contains less high-frequency amplitude than an ideal step when viewing its spectral profile. We used MATLAB's time-response functions to perform the math, given the previously computed rational function of the PDN and the rise-time of the step. The process is analogous to performing an inverse Fourier transform of the convolution of the step waveform with the PDN impedance. The result is the time-domain change in impedance, which is equivalent to the voltage deviation that the network creates given a 1A step-load with the appropriate rise-time. With the ability to modify the rise-time of the transient current excitation it is easy to re-compute the results, which may result in a different optimized set of capacitors as the frequency-content of the excitation changes.

With a computed step-response, we implemented the computation known as "reverse-pulse technique" [13] to calculate the worst-case peak to peak noise profile possible with this PDN. This technique assumes that the PDN behavior is linear and time-invariant (LTI), and the response is the same whether the step-current is positive or negative since each would induce an equal but opposite response. The approach does not take into account non-linear behavior that a VR may introduce, but in most cases the technique is sufficient to ascertain a worst-case response. The PDN impedance must have reasonable accuracy down to DC, and if the VR has active voltage positioning (AVP), also known as droop, or load-line [14], AVP also must be accounted for in the model. Incorporating the VR load-line is performed by adjusting the VR model included in the overall PDN model, or attached separately before optimization is started. The absolute DC voltage of the power rail is not important to predict the response, as the DC value would only appear as a simple voltage shift in the peak to peak response. Figure 5 illustrates a sample transient voltage response obtained from the capacitor optimizer application. The plot contains both undershoot and overshoot superimposed on the same graph, with DC shifting included. The x-axis scale is configured as logarithmic in order to observe clearly the high-frequency ringing.

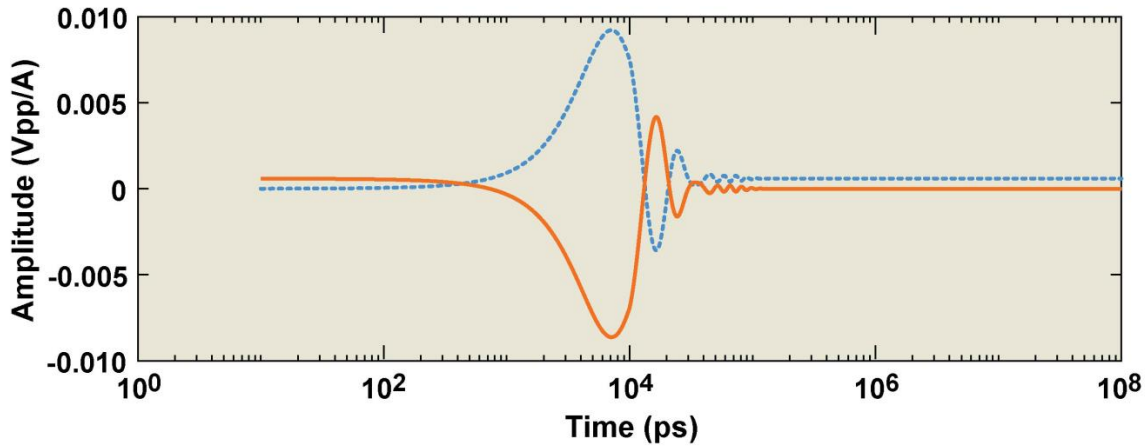

**Figure 5 - Transient response example (46592)**

The reverse-pulse technique computes the worst-case voltage response by summation of the peaks and valleys of the generated transient-response waveform. The process to sum the peaks and valleys is performed in reverse-chronological order. After adding the valleys and subtracting the peaks from the transient waveform, the two-sided transient response can be determined by doubling the result just computed, and finally subtracting the DC steady-state value. By subtracting the DC value, the VR load-line effects are included. The original work [13] on the reverse-pulse technique also developed the term "aperiodic resonant excitation" to describe the worst-case series of load transitions that would excite all the resonances in the PDN, and result in the worst-case behavior. A series of timed pulses is able to induce this behavior in a transient simulation. As an optional step, our application can output the worst-case pulse-train as a piecewise linear current source SPICE definition, to be used in a circuit simulator.

Figure 6 is an example of the aperiodic resonant excitation technique incorporated into our capacitor optimization process. The original response is on the left and the new optimized response on the right, using the transient response as the dominant scoring mechanism. These time-domain results are straight-forward and require less explanation of the optimized solution merits, as compared to a frequency-domain score based on target impedance. Even without computing the aperiodic resonant worst-case transient response, the PDN on the right exhibits less overall voltage deviation from a single pulse, which is not always the case when considering the summation of all the peaks and valleys of the response.

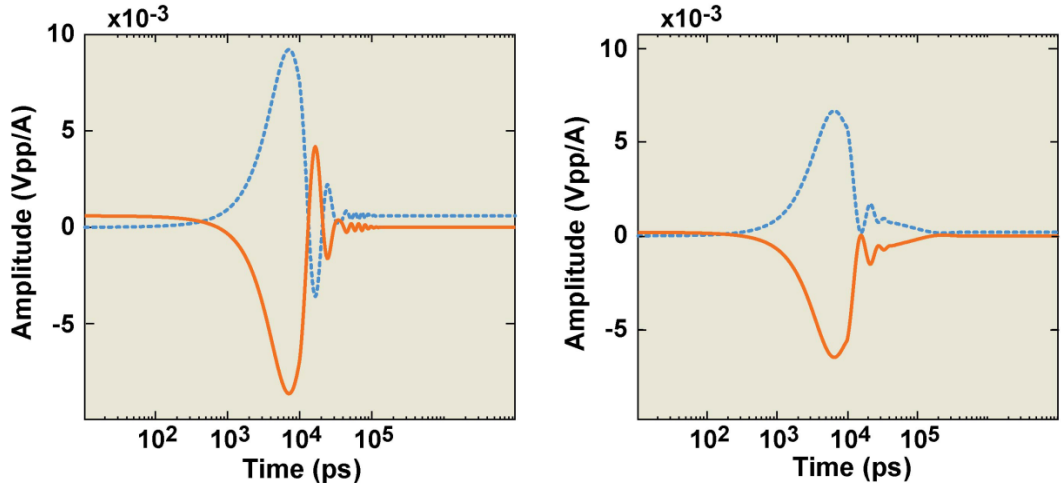

**Figure 6 – Aperiodic resonant excitation capacitor optimization, un-optimized (left) versus optimized (right) (46593)**

Including the transient scoring computation in the optimization process did not greatly impact our solution convergence time in most cases. The solution performance impact is primarily subject to the rational model order and the required time-steps, which are all adjustable.

# IV. Implementation Examples

An overview of using the capacitor optimization tool is depicted in Figure 7. We now demonstrate this process on first a simple PDN and then a more complex one.

## Model Preparation

The first step is to identify the capacitor reference designators to optimize for a given PCB PDN. The S-parameter model quickly becomes large and unwieldy if it contains too many ports and frequency points. For simple structures with tens of ports there is usually minimal difficulty extracting broadband models, but when hundreds of ports are involved the frequency points should be chosen strategically to minimize the file size and processing time. In order to speed up the port assignments and model extraction for larger PDNs, a few simple scripts are helpful. After we identify the list of capacitor sites to be optimized, we use a script to generate a text file containing the reference designators (ref-des) of each capacitor to be optimized. Most EM solvers have the ability to create and execute scripts directly within the simulator. We used the scripting environment to automate reading in the text file containing the capacitor list; and for each capacitor, generate a local port and de-activate the original capacitor element. After the script is executed, the model is ready to be set up for broadband extraction. To minimize execution time for the optimization algorithm, we constrained the resulting Touchstone models to be less than one gigabyte by adjusting the frequency sweep parameters. An interpolating frequency sweep is beneficial to ensure that peaks/valleys are captured while avoiding excess frequency points in flatter regions. Using a modern multi-core workstation the optimization time for large models could still be several hours. The majority of the time spent by our application is during the S-parameter matrix math calculations, while attaching capacitors using vectorized port reduction algorithms [10]. The solution time is CPU-intensive and additional techniques to optimize the port reduction processes may be beneficial for future speed improvement.

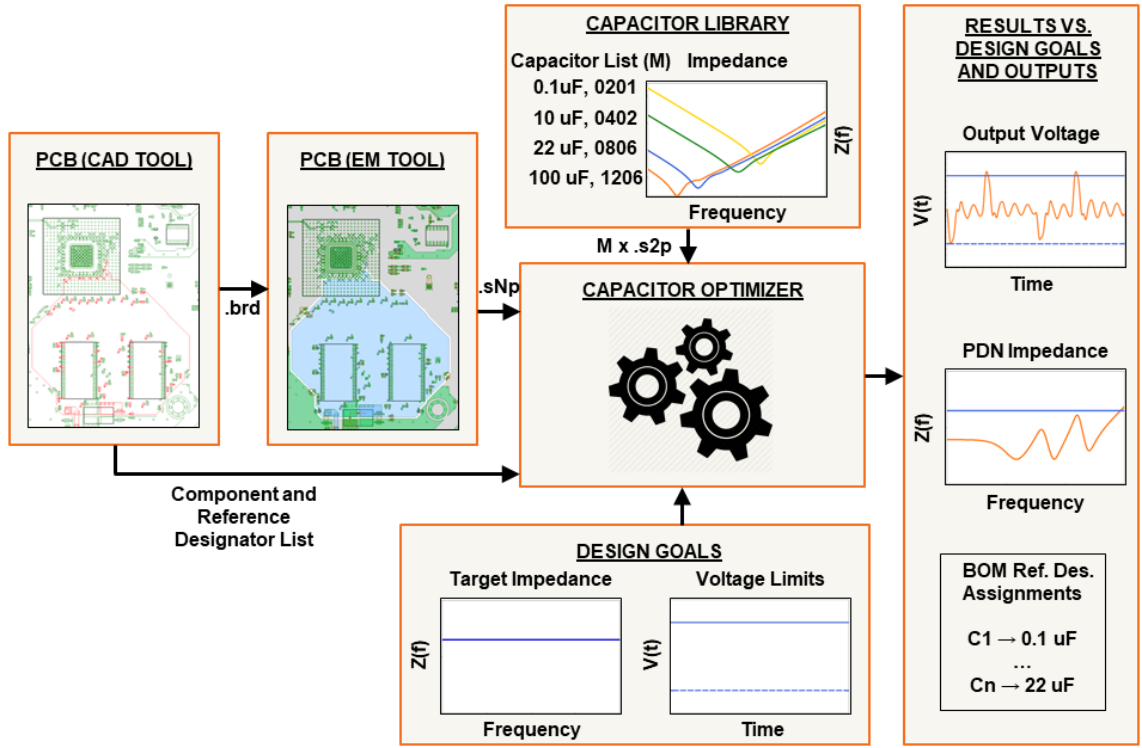

**Figure 7 - Overview of using the capacitor optimization tool on a PDN (46645)**

## Optimizing a Simple PDN

Our capacitor optimization application is invoked from within the MATLAB environment, where a new object is constructed with arguments defining:

- The name of the model to be optimized
- The die model to use, if any is desired
- The VR model to use, if any is desired
- The target impedance definition text-file
- The list of candidate capacitor models text file

After object instantiation the optimizer can be run, or the object can be manually manipulated to try capacitor attachment combinations, view plots, etc. For the purposes of this demonstration a simple PCB PDN was analyzed as shown in Figure 7. This example has 36 capacitors located across the power domain.

When the capacitor optimizer object is created, our code generates an initial capacitor population guess as illustrated in the top-left subplot of Figure 8, where the resulting PDN Z(f) and target impedance curve is displayed. The prominent features of this initial response are the relatively low impedance near 350 kHz and the high anti-resonance peak around 40 MHz due to the on-die capacitance interacting with the packaging inductance. The depicted response is not achieving the target anywhere in the range of 10 MHz to 100 MHz.

The optimizer is then configured to minimize the area above the target impedance curve while ignoring flatness or maximum impedance peaks. While the GA is running, our application presents a heads-up display that shows the convergence process, current population scores, and the generational score trends. The first optimization attempt focuses on minimizing the area between the target-Z curve and the PDN Z(f). The optimized Z(f) appears in the top-right subplot of Figure 8 as solution A. The results may be surprising when viewed on a logarithmic scale. Our optimizer was not able to find capacitors in the list of supplied components to meet the target impedance, so it focused on minimizing the integration of the impedance above the target impedance line. The area between the Z(f) response and the target-Z curve is integrated across frequency, and compared to the initial guess solution it reduced by about 25%. The region from 10 to 100 MHz is greatly reduced at the expense of the lower-frequency region from 10 kHz to 10 MHz. The overall flatness score degrades considerably with the additional anti-resonance peaking. The results may not be favorable but there is no hope of finding a solution remaining completely underneath the given target-Z profile. As a second attempt, we adjusted the optimization process to add the flatness score into the weighting, resulting in solution B, depicted in the bottom-left plot of Figure 8. Here, the Z(f) is more clearly improved, with a reduction in the integrated target-Z score, but also with reduced Q-factor peaking.

To concentrate specifically on reducing the anti-resonance peaking due to package L and on-die C, a maximum impedance delta score is next included in the weighting, displayed as solution C in Figure 8. Solution C demonstrates how trying to minimize the 40 MHz anti-resonance results in sacrifices in the lower-frequency band; however the flatness score did not degrade much. Interestingly, solution C involves a partial capacitor population in order to minimize the anti-resonance caused by the capacitor ESL.

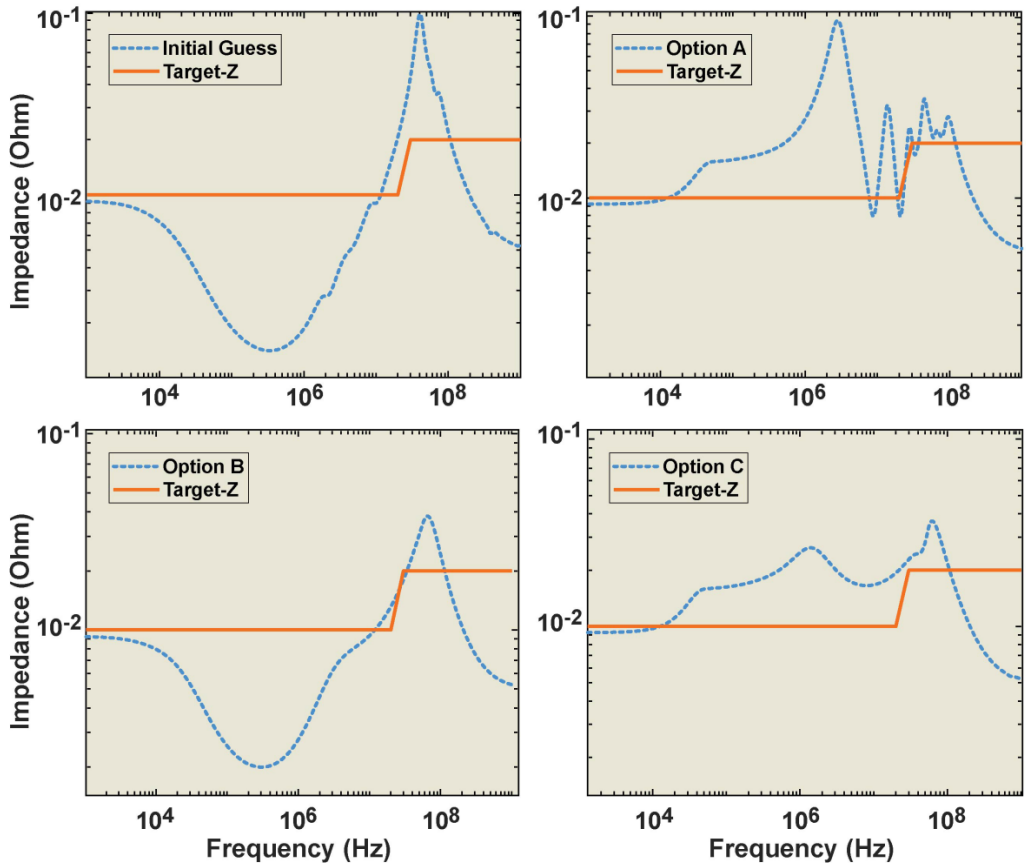

**Figure 8 - Target impedance optimization examples (46618)**

By now, four different capacitor attachment schemes have been presented by the optimizer in the frequency-domain, all using different methodologies and weighting. Which scheme is best? It may be visually obvious for simple cases, but there will often be doubt on which is the best predictor for low transient noise. Our application is able to generate the transient step-response waveform for each case, and those results are often easier to compare. In some cases the time-domain results are surprising when compared to the Z(f) plots.

For our final analysis, the transient optimization routine is used without any regard for the Z(f) scoring, with a pulse rise-time of 10 ns. Note that the source model includes a DC value of 10 mΩ, which indicates that there is significant resistance to the VR. The left panel of Figure 9 illustrates the starting capacitor population Z(f) profile versus the transient optimized Z(f) results. Comparing these two profiles shows the transient optimized results having flatter behavior, with higher impedance in the lower-frequency region and lower impedance in the higher-frequency anti-resonance band. If only presented with these Z(f) results, it may not be entirely obvious which solution would be best, but as further described in the right panel of Figure 9, the transient-optimized results are vastly superior to the original solution when using a 10 ns edge-rate for the step-response. The worst-case transient noise estimation demonstrates more than a 3x improvement with the transient-optimized PDN solution.

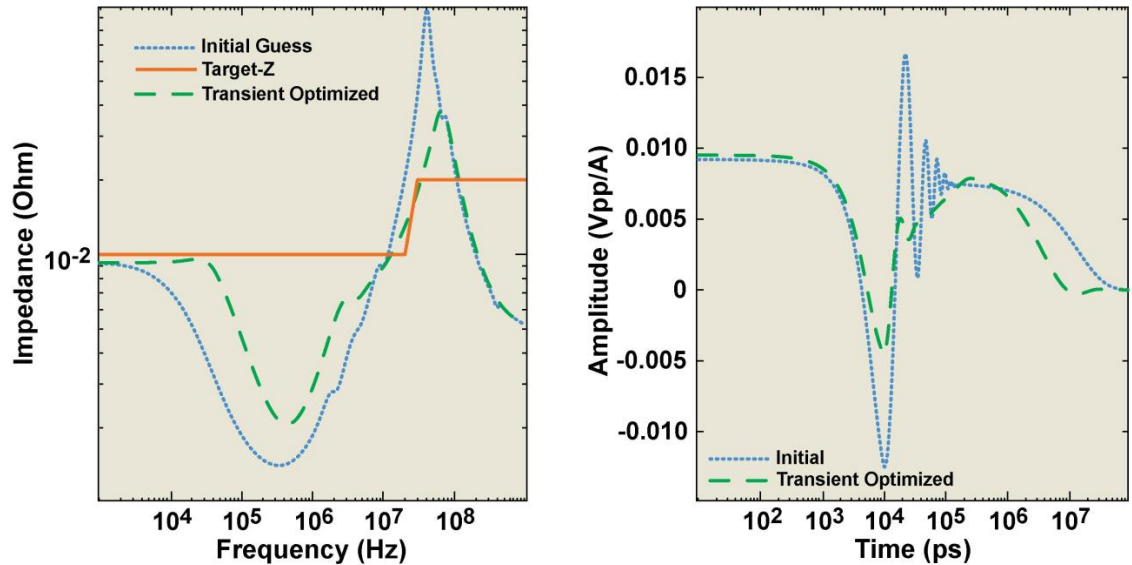

**Figure 9 - Initial guess versus transient optimized results, Z(f) (left) and transient (right) (46598)**

In Figure 10 the transient-optimized Z(f) is compared to the previous solution B. The similarity indicates for this PDN that the transient-optimized results and the solution B optimized results have converged to similar capacitor populations. Figure 10 shows the step-response comparisons, again demonstrating good correlation and well-matched results.

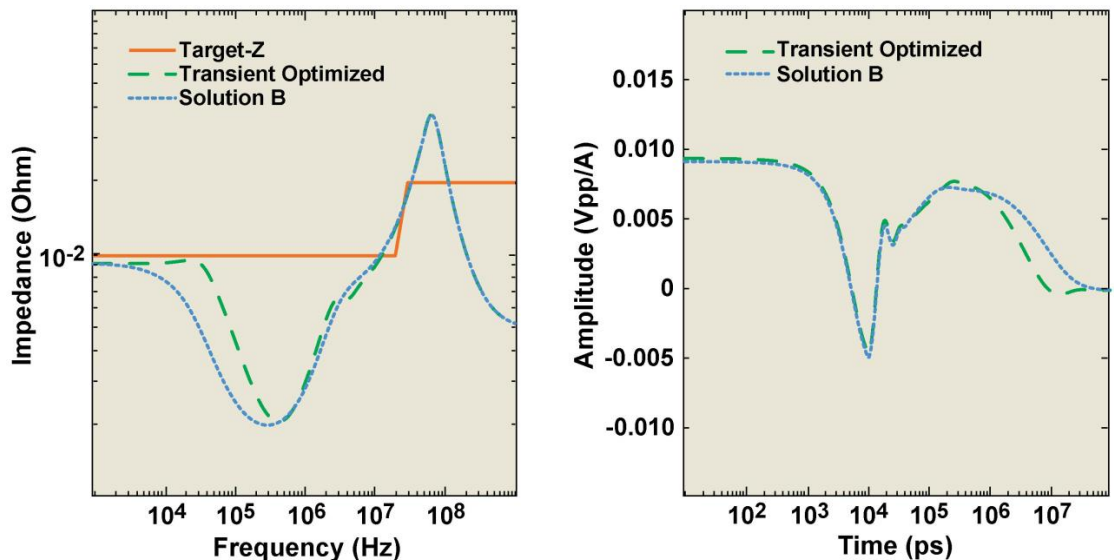

**Figure 10 – Transient versus Z(f) optimized solution B, Z(f) (left) and transient (right) (46599)**

For this PDN, through some trial and error, we found that a certain combination of our frequency-domain scoring criteria can achieve similar results to our transient optimizer. However, with the transient-optimized results a designer can avoid the complications of combining Z(f) scoring metrics and using engineering intuition or additional external simulations to confirm that the results are optimized. If the worst-case transient noise is what matters most, then the optimization should ideally be directly performed on that metric.

## Optimizing a Larger PDN

A new circuit board model was created in order to demonstrate scalability to large port-count models. Figure 11 presents an example structure created with 300 capacitor sites, an ASIC load (left) and a VR source (right). The S-parameter model was created containing 302 ports and roughly 30 frequency points per decade from DC to 500 MHz. The resulting Touchstone file was approximately 700 MB. A die model was attached in the capacitor optimization application in order to generate the anti-resonance between the die and the PDN inductance. Several candidate capacitor models were selected in the range of 10 nF to 22 µF. The target impedance was defined as 650 µΩ and rising at 30 MHz to a maximum of 3 mΩ at 100 MHz.

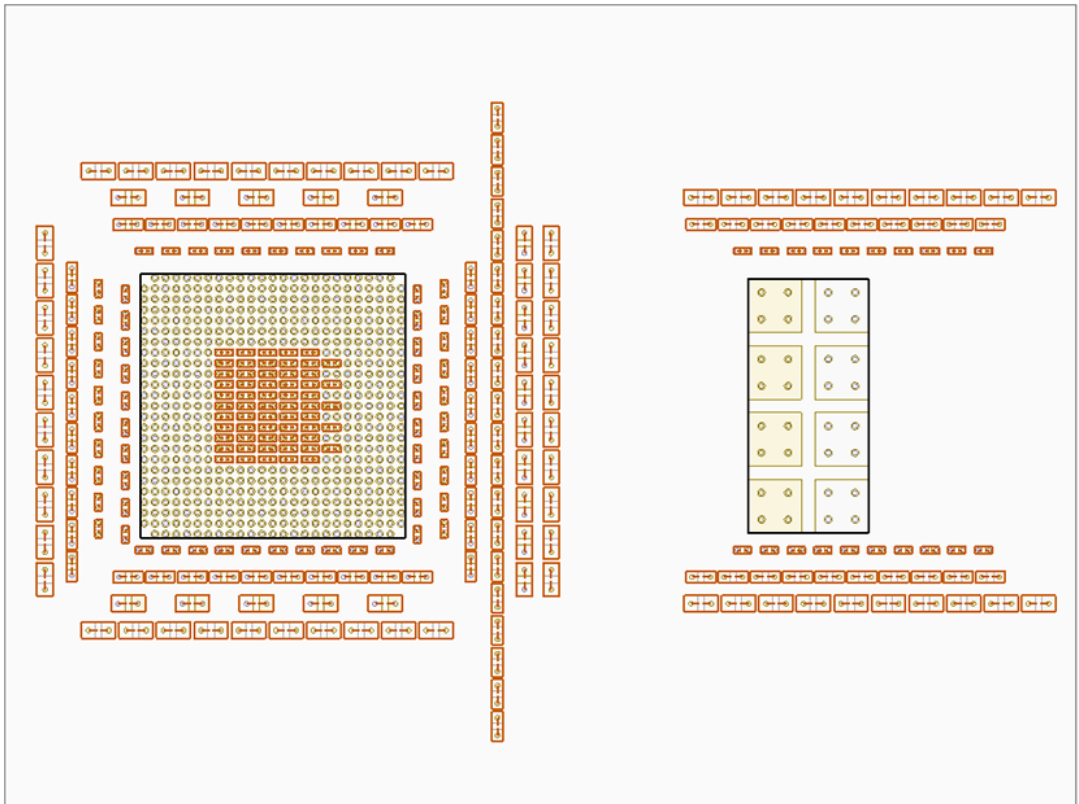

**Figure 11 - Large PDN circuit board model (46600)**

Figure 12 depicts a typical PDN solution using a target-impedance based capacitor optimization process. This solution maintains the target impedance by using a minimum set of capacitors, as illustrated in the left panel. Our transient optimizer routine was used to apply a 10 ns step-load waveform to the PDN response, with the resulting transient results appearing on the right of Figure 12. Calculating the worst-case aperiodic resonant excitation results in 1.12 mVpp/A. Even though this PDN does not appear to exceed the target impedance of 650 µΩ, when considering the aperiodic resonant excitation technique instead of the single-pulse load, the transient voltage requirements may not be met.

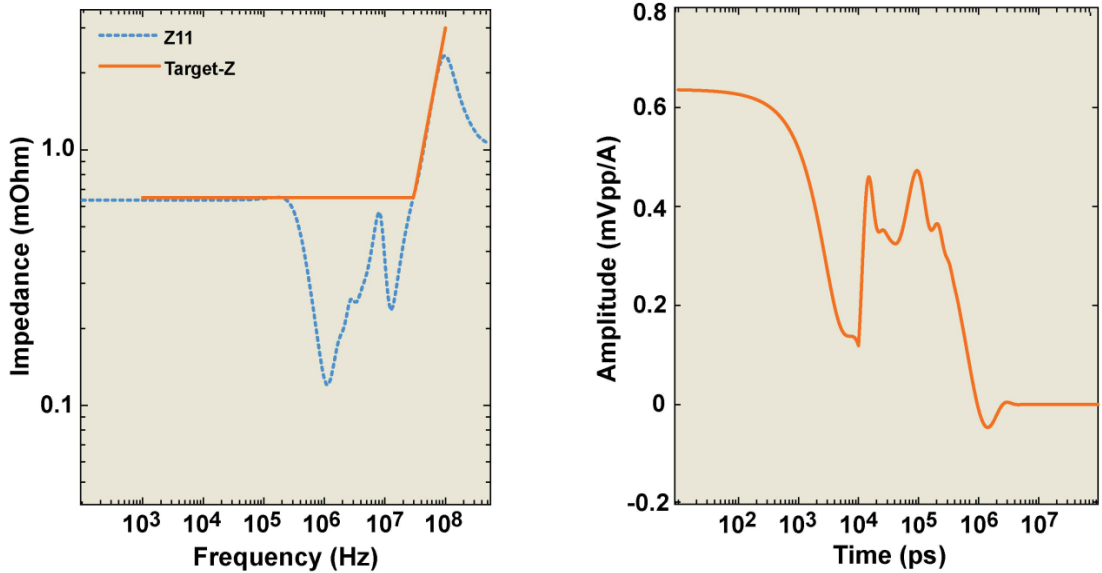

**Figure 12 - Achieving target impedance for a large PDN, Z(f) (left) and transient step-response (right) (46601)**

The same source PDN model was then run through the transient optimization routine and the resulting Z(f) and voltage response to a 10 ns step-load appears in Figure 13. The calculated worst-case aperiodic resonant excitation result is reduced to 0.658 mVpp/A, which is extremely close to meeting the intended 650 µΩ of network impedance, even in the worst-case pulse computation.

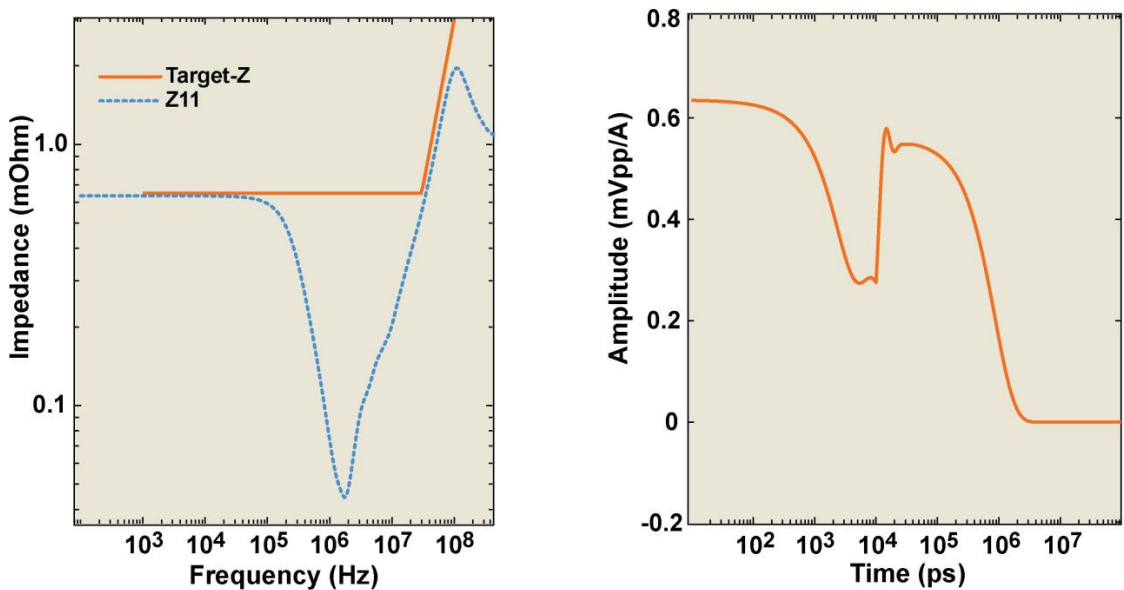

**Figure 13 - Large PDN that was transient optimized, Z(f) (left) and transient step-response (right) (46602)**

Our transient optimizer was successful in producing an optimized network that would have high confidence in meeting the intended target impedance goals. Computing the transient optimization results required slightly more than 5 hours on a modern high-performance computer. Note that the optimized result is not representative of the best possible network profile that could be achieved. The results are subject to the suitability of the capacitor models being used and the capacitor port locations within the PDN. Achieving a very flat Z(f) profile is challenging without a wide assortment of controlled ESR capacitors to dampen anti-resonances formed by capacitor interactions. Standard low-ESR high-frequency capacitors often generate anti-resonances that harm the overall worst-case transient response results. The availability of controlled-ESR capacitors is still limited despite their advantages in some cases [15].

## VI. Summary

A proposed methodology has been described and demonstrated to optimize PDN decoupling capacitors numerically using an end-to-end simulation tool. The details of our application construction were explained, and optimization methods were compared. The simple target impedance metric was discussed, including some of its shortcomings as an optimization metric. Many designs are challenged to achieve a target impedance goal, or tend to focus on optimizing for the best performance. This paper has illustrated multiple ways to assess performance. We have demonstrated that incorporating all of the best practices into a frequency-domain based optimization routine is challenging, especially when a simple pass/fail criteria is not adequate. We described the development and usage of a new worst-case transient analysis optimization method. Our optimizer application can be very easily extended to include new scoring algorithms to gauge their effectiveness. With the advent of the transient-response optimizer proposed in this paper, a natural extension could be to incorporate automatic optimization for the active VR load-line. Other useful features might include: netlist functionality in order to build PDN circuits, multi-objective optimization (Pareto optimization), supporting case-size specific capacitor model options, and providing an easier user front-end.